\documentclass[aps,prl,twocolumn,showpacs,amsmath,amssymb]{revtex4-1}

\usepackage{graphicx}
\usepackage{color}
\usepackage[breaklinks,colorlinks=true]{hyperref}

\newcommand{\bA}{\boldsymbol{A}}
\newcommand{\bB}{\boldsymbol{B}}
\newcommand{\bE}{\boldsymbol{E}}
\newcommand{\ble}{\boldsymbol{e}}
\newcommand{\bx}{\boldsymbol{x}}
\newcommand{\calD}{\mathcal{D}}
\newcommand{\calK}{\mathcal{K}}
\newcommand{\calP}{\mathcal{P}}

\DeclareMathOperator{\tr}{tr}
\DeclareMathOperator{\Tr}{Tr}

\DeclareMathOperator{\sech}{sech}

\begin{document}

\title{Spatially Assisted Schwinger Mechanism and Magnetic Catalysis}

\author{Patrick Copinger}
\affiliation{Department of Physics, The University of Tokyo, 
             7-3-1 Hongo, Bunkyo-ku, Tokyo 113-0033, Japan}

\author{Kenji Fukushima}
\affiliation{Department of Physics, The University of Tokyo, 
             7-3-1 Hongo, Bunkyo-ku, Tokyo 113-0033, Japan}

\begin{abstract}
 [This is an erratum applied version to Phys.\ Rev.\ Lett.\ {\bf 117},
   081603 (2016).  The conclusion is not changed, but we should have
   considered not decreasing but increasing spatial modulations in the
   magnetic profile.]
 Using the worldline formalism we compute an effective action for
 fermions under a temporally modulated electric field and a spatially
 modulated magnetic field.  It is known that the former leads to an
 enhanced Schwinger Mechanism, while we find that the latter can also
 result in enhanced particle production and even cause a
 reorganization of the vacuum to acquire a larger dynamical mass in
 equilibrium which spatially assists the Magnetic Catalysis.
\end{abstract}

\maketitle

\paragraph*{Introduction:}

The vacuum in quantum field theory is filled with fluctuation pairs of
particles and anti-particles and this fact causes various
unconventional phenomena such as the Casimir
force~\cite{Lamoreaux:2005gf}, the Schwinger
Mechanism~\cite{PhysRev.82.664,*sauter,*Heisenberg:1935qt} (see
Ref.~\cite{Dunne:2004nc} for a review), the spontaneous symmetry
breaking, and so on.  Among
the above examples the Schwinger Mechanism still awaits an
experimental confirmation.  The idea is that particle and
anti-particle (i.e.\ $e^-$ and $e^+$) pairs should be created from the
vacuum driven by a strong external electric field $\bE$, which is
analogous to the Landau-Zener effect in materials.  The particle
production rate $w$ in the Schwinger Mechanism is, however,
exponentially suppressed as $w\sim e^{-\pi m_e^2/(eE)}$.  The critical
electric field, $eE_c\sim \pi m_e^2$, is too strong to be reachable in
current laboratory experiments and the detection of Schwinger pair production
with strong laser fields has remained elusive (see
Ref.~\cite{Dunne:2008kc,*Gies:2008wv} for 
reviews).

It is evident from such an exponential form that the electron mass
could be given an interpretation as an ``activation energy'' in atomic
ionization, and it has been concluded that the ionization is favored
more by time-dependent $\bE(t)$ in the pioneering work by
Keldysh~\cite{keldysh} but less by space-dependent
$\bE(\bx)$~\cite{Nikishov1970346}.  This observation
suggests that there might be an optimal profile of $\bE(\bx,t)$ that
enhances the pair production rate in the Schwinger Mechanism.  Indeed,
the ``Dynamically Assisted Schwinger Mechanism'' has been recognized
by Sch\"{u}tzhold, Gies, and Dunne~\cite{PhysRevLett.101.130404} who
first successfully quantified how much a time-dependent perturbation
in the electric field can push the critical field strength down.
Numerical simulations with spacetime-dependent $\bE(\bx,t)$
have been performed later~\cite{Hebenstreit:2011wk}.

A natural question would be: What if an external magnetic field $\bB$
co-exists?  For homogeneous and constant $\bE$ and $\bB$ the
Euler-Heisenberg Effective Hamiltonian already contains both effects
of $\bE$ and $\bB$.  In particular, if there are parallel components,
i.e.\ $\bE\cdot\bB\neq 0$, the particle production rate $w$ has an
anomalous origin associated with the non-conservation of axial
charge or the axial
anomaly~\cite{Iwazaki:2009bg,PhysRevLett.104.212001,*Fukushima:2015tza}.
Then, an optimal profile of $\bB(\bx,t)$ might further increase $w$.
In the present work we will show that an increasing or positive curvature
space-dependent perturbation
in the magnetic field decreases the effective mass of fermions and
thus the critical field strength, which we call the ``Spatially Assisted
Schwinger Mechanism.''

Such a situation with space-time modulated $\bE$ and $\bB$ would be
realistic in nucleus-nucleus collisions at high energy as conducted in
Relativistic Heavy-Ion Collider (RHIC) and in Large Hadron Collider
(LHC).  In each collision event thousands of particles (mostly pions)
are created and a non-Abelian generalized Schwinger Mechanism should
underlie the particle
production~\cite{PhysRevD.21.1095,*Ambjorn1983340,*Gyulassy1985157,PhysRevC.75.044903,*Tanji20102018}.

Besides, a strong $\bB$ is expected in non-central collisions, which
has inspired enormous activities dedicated to fermionic matter at
strong constant $\bB$ with lattice numerical
simulations~\cite{Bali:2014kia} and phenomenological model
studies~\cite{Gatto:2012sp}.  One of the most remarkable implications
from the presence of such strong $\bB$ is the Chiral Magnetic Effect~\cite{PhysRevD.78.074033,Kharzeev2010205};  C-
and CP-odd (chromo-electromagnetic) backgrounds supply the system with
a finite net chirality and an electric current is then induced along
$\bB$.  The Chiral Magnetic Effect is an interdisciplinary subject
ranging over nuclear physics, astrophysics, and especially condensed
matter physics in which Dirac/Weyl semimetals and graphene could
provide us with clean environments under better experimental control
(also applicable for the Schwinger
Mechanism)~\cite{PhysRevB.92.085122,*PhysRevD.78.096009}.  So far, most
of theoretical works on the Chiral Magnetic Effect are limited to
homogeneous and constant $\bE$ and $\bB$ (see
Ref.~\cite{PhysRevD.86.085029} for an exception), though the
nucleus-nucleus collision generates electromagnetic fields with large
spatial modulations~\cite{1674-1137-39-10-104105}.

So far, we have focused on the particle production in real time, but
our finding of the spatially assisting mechanism due to an effective
mass shift implies that the true vacuum should be modified in
equilibrium.  It is widely known that
homogeneous and constant $\bB$ leads to the Magnetic Catalysis, i.e.\
enhancement of dynamical symmetry breaking or catalyzing the fermion
condensate~\cite{Klimenko:1990rh,*Klimenko:1992ch,*PhysRevLett.73.3499,*Shovkovy:2012zn}.
Because the mass shift shows completely the same pattern as the Chiral
Gap Effect in curved
space~\cite{PhysRevLett.113.091102,*PhysRevLett.114.181601}, we can
draw the same conclusion from our calculation;  spatially modulated 
$\bB$ increases the dynamical mass by the reorganization of the vacuum
structure.
To the best of our knowledge, the present work is the very first
demonstration that spatially modulated $\bB$ should assist the
Magnetic Catalysis as well as the Schwinger Mechanism.

For actual calculations we employ the worldline path-integral
formalism~\cite{Schubert200173}, which is suited to handle pair
particle production with general background fields.  In this work we
will not take account of back-reaction, that is, effects of additional
electromagnetic fields sourced by produced charged particles, since we
are interested in the production rate only for a given configuration
and not the whole temporal evolutions.
\vspace{0.3em}

\paragraph*{Worldline Formalism:}

Our starting point is the (4+1)-dimensional Euclidean worldline
effective action $\Gamma_{\rm E}$ for a fermion with mass $m$ under a
given gauge configuration $A_\mu(x)$, which is generally expressed
as~\cite{Schubert200173}
\begin{align} 
  &\Gamma_{\rm E}[A] = \int_0^\infty \frac{dT}{T}\, e^{-m^2 T}
  \oint \calD x \notag\\
  &\quad\times \exp\biggl\{-\int_0^T d\tau\,\biggl[
   \frac{1}{4}\Bigl(\frac{dx}{d\tau}\Bigr)^2
   + ieA\cdot \frac{dx}{d\tau} \biggr]\biggr\}\,\Phi[A] \;,
\label{eq:world1} \\
 &\Phi[A] = -\frac{1}{2}\tr \calP \exp\biggl(
  \frac{ie}{2} \int^T_0 d\tau\, \sigma_{\mu\nu}F_{\mu\nu} \biggr) \;,
\label{eq:spin}
\end{align}
where $\Phi[A]$ represents the spin factor with
$\sigma_{\mu\nu}\equiv\frac{1}{2}[\gamma_\mu,\gamma_\nu]$.  The
Euclidean field strength tensor components are
$F_{ij}=\epsilon_{ijk}B_k$ and $F_{4i}=-iE_{i}$ in terms of the
physical (Minkowskian) electromagnetic fields.  In the worldline
formalism the auxiliary coordinate variables should satisfy the
periodic boundary condition; $x_\mu(0) = x_\mu(T)$.  It is important
to note that the $m$-dependence appears only through $e^{-m^2 T}$
which makes the $T$-integration converge as long as $m^2$ is
positive.

In the present study we specifically consider a situation with the
$x_4$-dependent $\bE$ and the $x_{1,2}$-dependent $\bB$ fields
parallel to each other along the $x_3$-axis;
i.e.\ $\bE=E(A_4(x_{3,4}),A_3(x_{3,4}))\ble_3$ and
$\bB=B(A_1(x_{1,2}),A_2(x_{1,2}))\ble_3$.  In this special
but topologically non-trivial situation with $\bE\cdot\bB\neq 0$, the
electric and the magnetic parts are separable (but coupled through the
$T$-integration) and we can express the effective action as
$\Gamma_{\rm E}=-2\int^\infty_0 \frac{dT}{T} e^{-m^2T}
\mathcal{K}_E \mathcal{K}_B$, where
\begin{align} 
 &\calK_E(T;A_3,A_4) \equiv
  \oint\calD x_3 \calD x_4\, \cos\biggl[\int_0^T d\tau\,
  eE(x) \biggr] \notag \\
  &\qquad\times \exp\biggl\{-\int_0^T d\tau\, \biggl[\frac{1}{4}
    \Bigl(\frac{dx_3}{d\tau}\Bigr)^2+\frac{1}{4}\Bigl(
    \frac{dx_4}{d\tau}\Bigr)^2 \notag\\
  &\qquad\qquad\qquad\qquad\qquad + ieA_3 \frac{dx_3}{d\tau}
    + ieA_4 \frac{dx_4}{d\tau} \biggr]\biggr\}\;.
\label{eq:lage} \\
 &\calK_B(T;A_1,A_2) \equiv
  \oint\calD x_1 \calD x_2\, \cosh\biggl[\int_0^T d\tau\,
  eB(x) \biggr] \notag \\
  &\qquad\times \exp\biggl\{-\int_0^T d\tau\, \biggl[\frac{1}{4}
    \Bigl(\frac{dx_1}{d\tau}\Bigr)^2+\frac{1}{4}\Bigl(
    \frac{dx_2}{d\tau}\Bigr)^2 \notag\\
    &\qquad\qquad\qquad\qquad\qquad + ieA_1 \frac{dx_1}{d\tau}
    + ieA_2 \frac{dx_2}{d\tau} \biggr]\biggr\}\;.
\label{eq:lagb}
\end{align}
The theoretical treatment of $\calK_E$ has been well investigated:
As long as $\calK_B$ does not produce any exponential factor in terms
of $T$, we can perform the $T$-integration approximately with the
value at the stationary point $T^\ast(x_{3,4})$.  Then, we can
evaluate the $x_{3,4}$-integration with the worldline instantons
$x^{\rm cl}_{3,4}(\tau)$ as solutions of the equations of
motion~\cite{AFFLECK1982509,*PhysRevD.72.105004} including Gaussian
fluctuations around them to reproduce the prefactor.  Finally we can
just replace the magnetic part with
$\calK_B(T^\ast(x^{\rm cl}_{3,4});A_1,A_2)$. 
In this way, for constant $\bE\parallel \bB$, pretty straightforward
calculations yield the correct answer;
$\Gamma_{\rm E}=-[e^2 EB/2(2\pi)^2]\coth(\pi B/E)\exp(-\pi m^2/eE)$ by
picking up the $n=1$ instanton contribution.  This can also be
interpreted as a particle production rate in Minkowskian spacetime as
\begin{equation}
 w=2{\rm Im}\Gamma_{\rm M}=-2{\rm Re}\Gamma_{\rm E}
 = \frac{e^2 EB}{(2\pi)^2}\coth\Bigl(\frac{\pi B}{E}\Bigr)
    e^{-\pi m^2/eE}\;,
\label{eq:w_EB}
\end{equation}
which happens to coincide with the exact answer from the
Euler-Heisenberg effective Lagrangian.  What we will show below is
that, with spatially modulated magnetic fields, $\calK_B$ may give
rise to an exponential factor and this effectively changes $m^2$.
\vspace{0.3em}

\paragraph*{Magnetic Part with Spatial Modulation:}

Below we report our calculations.
Let us first consider an example with the following
(Sauter-type) profile of the magnetic
field, the Dirac equation with which can be
solved~\cite{PhysRevD.52.R3163,*Dunne1998322};
\begin{equation}
 \bB(x) = B \sech^2(\kappa x_1)\,\ble_3\;.
\label{eq:Bsech}
\end{equation}
It would be more convenient to translate the functional integral into
the canonical quantized representation with the corresponding
Hamiltonian.  Then, we can re-express $\calK_B$ into the following
form as
\begin{equation}
 \calK_B = \sum_\pm \frac{1}{2}\Tr\exp(-\hat{H}_B^\pm T)\;,
\end{equation}
where the magnetic Hamiltonian is defined as
\begin{equation}
 \hat{H}_B^\pm \equiv -\partial_1^2 - \biggl[\partial_2
  + \frac{ieB}{\kappa}\tanh(\kappa x_1)\biggr]^2
  \pm eB\sech^2(\kappa x_1)\;.
\label{eq:magham}
\end{equation}
In the same way as in
  Ref.~\cite{PhysRevD.52.R3163,*Dunne1998322} we can find the
  wave-functions to diagonalize the above Hamiltonian,
and from them we can construct the eigenvalue spectrum explicitly.
Specifically, the kernel takes a form of
\begin{equation}
 \Tr e^{-\hat{H}_B^\pm T} = -\frac{1}{2\pi i}\int\frac{dp_2}{2\pi}
  \int_{C} d\lambda\, e^{-\lambda T} g^\pm(\lambda)\;.
\end{equation}
Here, $g^\pm(\lambda)$ is an integrated resolvent made from the
wave-functions and after some calculations we find,
\begin{equation}
  g^\pm(\lambda) = \tilde{g}(p_2, eB, \kappa, \lambda)
  \sum_{j=\pm} \psi\Bigl(\frac{1}{2}+j
  \Bigl|\frac{1}{2}\pm\frac{eB}{\kappa^2}\Bigr|
  +\epsilon_-+\epsilon_+\Bigr)\;,
\end{equation}
where we defined; $\tilde{g}(p_2, eB, \kappa, \lambda)
\equiv\kappa^4 (\epsilon_-+\epsilon_+)^3/[(p_2\,eB)^2
  -\kappa^6(\epsilon_-+\epsilon_+)^4]$ and the dimensionless
dispersion relations are
$\epsilon_\pm(\lambda)\equiv(2\kappa)^{-1}\sqrt{(p_2\pm eB/\kappa)^2-\lambda}$,
and $\psi(x)$ represents the digamma function.  What is necessary for
our present purpose is to locate the poles of $g^\pm(\lambda)$ and they
are identified from the properties of the $j=-1$ digamma function.
After some procedures we have discovered the explicit form of the
eigenvalue spectrum as
\begin{align}
 \lambda_n^\pm &= p_2^2 \biggl[ 1-\frac{(eB)^2}
  {(\kappa^2\tilde{n} - |\kappa^2/2 \pm eB|)^2} \biggr] \mp eB \notag \\
 &\qquad\qquad - \biggl( \kappa^2 \tilde{n}^2 - 2\tilde{n}
  \Bigl|\frac{\kappa^2}{2} \pm eB\Bigr| + \frac{\kappa^2}{4}
  \biggr)\;,
\end{align}
where a half integer $\tilde{n}\equiv n+1/2$ ranges with
$n \in [ 0, |\frac{1}{2} \pm \frac{eB}{\kappa^2}|-\frac{1}{2}
-\sqrt{\frac{p_2 eB}{\kappa^3}})$ for $\lambda_n^\pm$.
We can significantly simplify the above expression for
$eB>\kappa^2/2$, which is the case for our interested situation with
small inhomogeneity.  Then,
\begin{equation}
  \lambda_n^\pm = \biggl[1-\frac{p_2^2 \kappa^2}{(eB-\kappa^2 n^\pm)^2}
    \biggr]n^\pm (2eB-\kappa^2 n^\pm)\;,
  \label{eq:eigen}
\end{equation}
where $n^+=n$ and $n^-=n+1$.
As we described before, in the presence of the electric field, the
$T$-integral can be approximately evaluated at the stationary point
and the smallest $\lambda_n^\pm$ would dominate.  Thus, we pick up the
contributions from $n=0$ only, namely, $\lambda_0^+=0$.
Interestingly, the second smallest eigenvalue appears
from the largest $p_2$, which is set by
the condition, $n\ge 0$, leading to
$\lambda_1^+\bigr|_{p_2=(eB-\kappa^2)/(\kappa eB)}
=\lambda_0^-\bigr|_{p_2=(eB-\kappa^2)/(\kappa eB)}\simeq 4\kappa^2$.
Thus, the energy gap is characterized by not $eB$ but $\kappa^2$.

Instead of descreasing behavior for small $x_1$ in the Sauter-type
shape, let us consider increasing behavior.  One may then expect to
have a negative $\lambda_n^\pm \propto -\kappa^2$.
Actually, in an inhomogeneous magnetic field setup -- a
magnetic flux tube with finite radius -- a negative-energy eigenstate
was found in Ref.~\cite{PhysRevD.60.105019}.  In view of the effective
action~\eqref{eq:world1}, such an exponential damping factor of
$e^{\kappa^2 T}$ should be interpreted as an effective  mass shift
as
\begin{equation}
  m^2 \to \tilde{m}^2 = m^2 - \kappa^2 \;.
\label{eq:shift}
\end{equation}
We note that in this work we call $\tilde{m}$ an
\textit{effective mass}, while we use a similar terminology, a
\textit{dynamical mass}, to mean a mass determined by the effective
potential.  We strictly distinguish them.

Now, let us confirm the above expectation by explicit calculations.
For the increasing magnetic profile, for small enough $\kappa^2$, we
can take the magnetic field and the associated vector potential as
\begin{align}
 \bB(x) &= \Bigl[B+B\frac{\kappa^2}{2}(x_1^2 + x_2^2)\Bigr]\ble_3\;,
\label{magneticfield} \\
 \bA(x) &= \frac{B}{2}(x_1\ble_2 - x_2\ble_1)
  -\frac{B\kappa^2}{6}(x_1^3\ble_2-x_2^3\ble_1)\;.
\label{magneticfieldgauge}
\end{align}
Because we are interested in the most dominant exponential factor for
large $T$, we do not have to solve the full eigenvalue spectrum but
can just compute the ground state energy or the vacuum energy.  We can
use the standard diagrammatic technique to obtain the vacuum energy as
a power series in $\kappa^2$ from the Lagrangian,
\begin{align}
  L &= \frac{(\frac{dx_1}{d\tau})^2+(\frac{dx_2}{d\tau})^2}{4}
    +\frac{ieB}{2}\Bigl( -x_2\frac{dx_1}{d\tau}+x_1\frac{dx_2}{d\tau}\Bigr)
    \pm eB \notag\\
    &~ \mp \frac{eB\kappa^2}{2}(x_1^2+x_2^2) + \frac{ieB\kappa^2}{6}
    \Bigl(x_2^3\frac{dx_1}{d\tau}-x_1^3\frac{dx_2}{d\tau}\Bigr)\;.
    \label{eq:lag}
\end{align}
To the first order in $\kappa^2$, the first term in the second line of
Eq.~\eqref{eq:lag} makes a contribution of one-loop vacuum graph,
which yields $\pm \frac{1}{2}\kappa^2$.  The second term makes a
contribution of two-loop vacuum graph, which yields
$-\frac{1}{2}\kappa^2$.  The sum amounts to the smallest energy of
$-\frac{1}{2}\kappa^2-\frac{1}{2}\kappa^2=-\kappa^2$, which confirms
the mass shift of Eq.~\eqref{eq:shift}.
We note that, for the Sauter-type decreasing profile, the sign of
$\kappa^2$ is opposite and then the smallest energy would be
$-\frac{1}{2}\kappa^2+\frac{1}{2}\kappa^2=0$ which is consistent with
$\lambda_0^+=0$ in Eq.~\eqref{eq:eigen}.
\vspace{0.3em}

\paragraph*{Spatially Assisted Schwinger Mechanism:}

The evaluation of $\calK_E$ is a well-known computation and we quickly
look over key equations here.  The vector potential is supposed to
have both terms of a constant electric field and a small perturbation
(with $e\varepsilon\omega \ll eE$) as
\begin{equation}
 A_3(x_4) = -i E x_4 -  i  \varepsilon  \tan( \omega x_4)\;.
\label{eq:dyngauge}
\end{equation}
Hereafter we use a rescaled proper time; $\tau=T u$.  After this
rescaling, the $T$-dependence in $\calK_E$ appears only in the
coefficient of $\dot{x}_3^2+\dot{x}_4^2$, where
$\dot{x}_{3,4}\equiv dx_{3,4}/du$. Because of the mass shift, the
stationary point~\cite{AFFLECK1982509,PhysRevD.72.105004} is modified
as
$T^\ast = \sqrt{\int_0^1 du\,(\dot{x}_3^2+\dot{x}_4^2)}/(2\tilde{m})$.
Then, regardless of the concrete choice of the gauge potential, we can
understand that $\dot{x}_3^2+\dot{x}_4^2$ is independent of $u$ (or
$\tau$) from the equations of motion, namely,
$\dot{x}_3^2+\dot{x}_4^2=C_n^2$ and $C_n=2n\pi m/(eE)$ is found where
$n$ refers to the instanton number.  By taking the $n=1$ contribution
we can get $\calK_E$ from the corresponding instanton action, and
we can eventually get
the particle production rate with the dynamically assisting $E$ and
the spatially assisting $B$ as
\begin{equation}
  w(\omega,\kappa) = \frac{e^2 EB}{(2\pi)^2}
  \coth\biggl(\frac{\pi B}{E}\biggr)\,
  e^{-S_{\rm inst}(\omega,\kappa)}\;,
\label{eq:w}
\end{equation}
where, for the modified Keldysh parameter
$\tilde{\gamma}\equiv \tilde{m}\omega/(eE)\ge \pi/2$, the instanton
action reads,
\begin{equation}
  S_{\rm inst}(\omega,\kappa) = \frac{2\tilde{m}^2}{eE}\Biggl[
    \arcsin\Bigl(\frac{\pi}{2\tilde{\gamma}}\Bigr)
    +\Bigl(\frac{\pi}{2\tilde{\gamma}}\Bigr)
      \sqrt{1-\Bigl(\frac{\pi}{2\tilde{\gamma}}\Bigr)} \Biggr]\;.
\end{equation}

\begin{figure}
  \includegraphics[width=0.95\columnwidth]{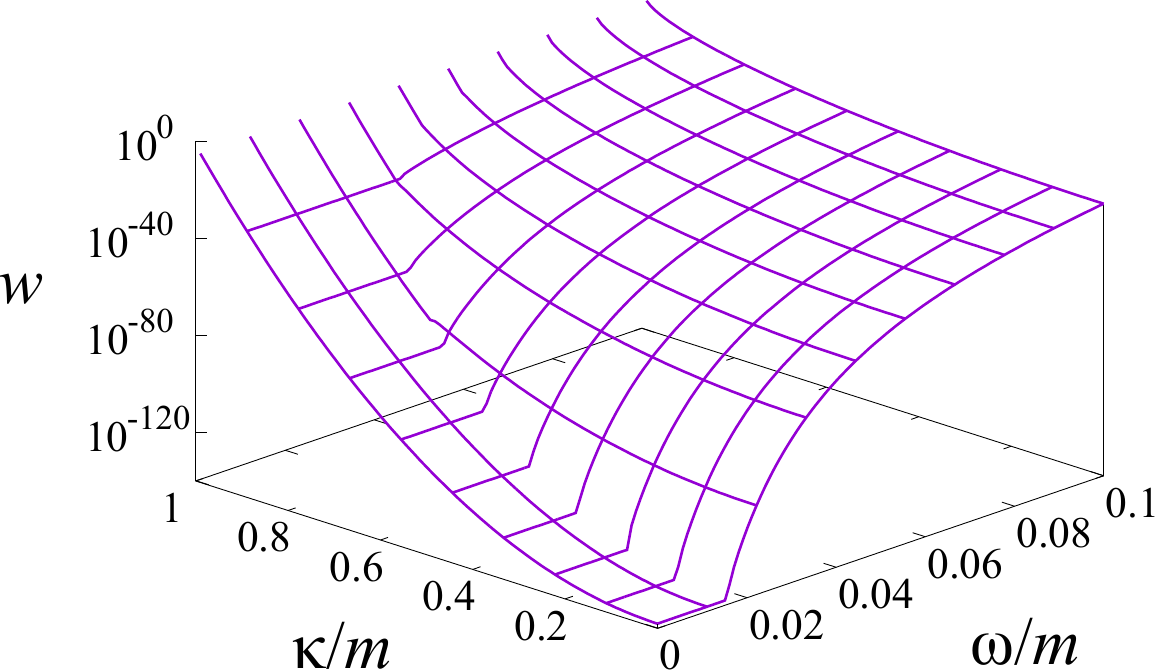}
  \caption{Pair production rate with the dynamically assisting $E$
    with frequency $\omega$ and the spatially assisting $B$ with
    wave-number $\kappa$ in unit of the particle mass $m$, where we
    chose $eE=eB=10^{-2}m^2$.}
  \label{fig:rate}
\end{figure}

\begin{figure}
  \includegraphics[width=0.95\columnwidth]{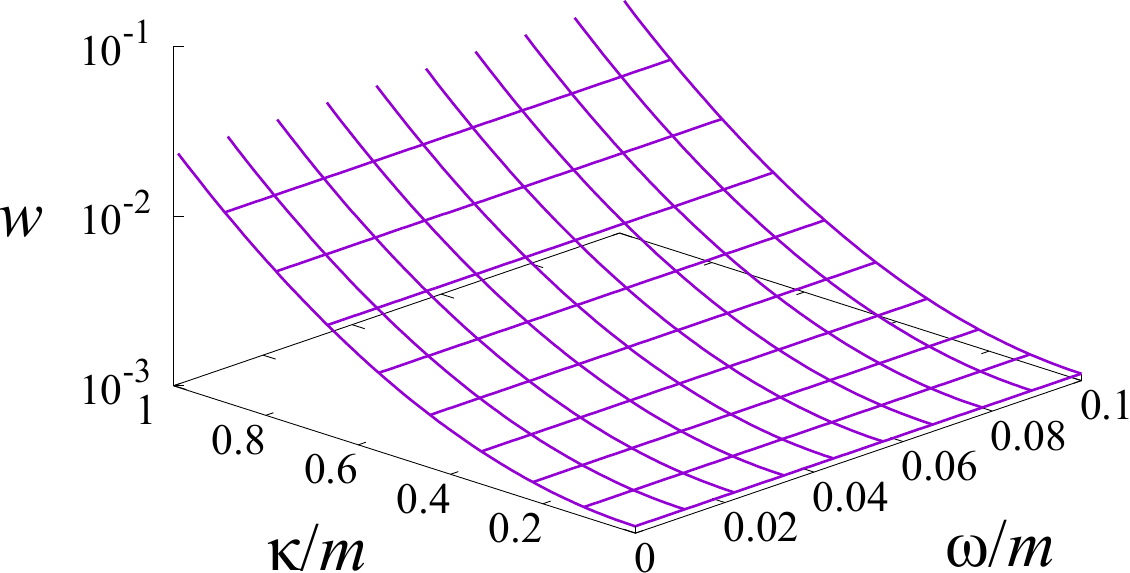}
  \caption{The same as Fig.~\ref{fig:rate}, where we chose
    $eE=eB=m^2$.}
  \label{fig:rate2}
\end{figure}

To have more insight, we make plots in Figs.~\ref{fig:rate} and
\ref{fig:rate2} to show $w(\omega,\kappa)$ in Eq.~\eqref{eq:w} as a
function of $\omega$ and $\kappa$ for different $E$ and $B$.
It should be noted that in the absence of any electric field we cannot
see pair production coming solely from the spatially inhomogeneous
magnetic field.  Strictly speaking, our expanded result makes sense
under the condition, $\kappa^2 \ll eB$, and if $\kappa\sim m$ and
$eE \ll m^2$ as is the case in the laser experiment, we require
$eB\gg eE$.  However, we should be aware that we can have an
unexpanded result for the Sauter-type potential~\eqref{eq:Bsech}
(which will be reported in more details in a follow-up) and
$\kappa^2 \ll eB$ is not mandatory for the Spatially Assisted
Schwinger Mechanism.  Here, to discuss qualitative characters, we
adopt $eE=eB$ for Figs.~\ref{fig:rate} and \ref{fig:rate2}.  From
Fig.~\ref{fig:rate} with $eE=eB< m^2$ (which would be more relevant to
the laser experiment) we see that the dynamically assisting effect has
a larger slope at small $\omega$ but gets saturated soon with
increasing $\omega$, while the spatially assisting effect has rather
opposite behavior.  In contrast to this, as seen in
Fig.~\ref{fig:rate2} with $eE=eB\gtrsim m^2$
(which would be more relevant to the nucleus collision
experiment), $\tilde{\gamma}$ is smaller, and the dynamically
assisting effect becomes minor, but the spatially assisting effect
remains prominent.

  We note that standing-waves of hard X-rays could in
  principle realize $\kappa\sim m_e=511\;\text{keV}$, and furthermore,
  in nucleus collision experiment $\kappa$ originates from the
  chromo-magnetic fields whose typical scale is given by the
  saturation scale $Q_s\sim 2\;\text{GeV}$ (see
  Ref.~\cite{Gelis:2015gza,*Fukushima:2016xgg} for recent reviews)
  that is thousands times greater than masses of quarks and gluons.
\vspace{0.3em}

\paragraph*{Spatially Assisted Magnetic Catalysis:}

An interesting question is; what happens if $\kappa^2 > m^2$ or
$\tilde{m}^2<0$?  Actually, as mentioned above, this is the case in
the nucleus collision.
Also in the laser experiment, such a situation could be
realized by means of the Weyl/Dirac semimetals in which
  fermions are nearly gapless and $\kappa^2 > m^2$ is easily
  achieved.
It would be important
to note that $\tilde{m}^2<0$ is not an artifact of our approximation;
in principle we could think of a massless theory
for which an infinitesimal $\kappa$ would realize $\tilde{m}^2<0$.
Then, we immediately notice that the $T$-integration no longer
converges.  What is the remedy for this apparent
breakdown?

We shall point out that the naive calculation of $w$ with a fixed $m$
should hold only transiently once we take account of interaction
effects, because we then consider the particle production problem on a
wrong vacuum.  To illustrate this, we could have utilized
  an interacting fermionic model such as the Nambu--Jona-Lasinio model,
  but here let us give a general and thus robust argument with the
  Ginzburg-Landau type effective potential.
For interacting fermions $m$ should be promoted
to be an in-medium mass or dynamical mass $M$, so that $M$ should be
self-consistently determined by the gap equation.  Let us suppose that
the gap equation follows from the minimum of an effective potential
whose Ginzburg-Landau expansion is $V(M^2) = a(M^2 - m_0^2)^2$ with
$a>0$ and $m_0 > m$.  Then, naturally, the vacuum (ground state)
should favor $M^2=m_0^2$ to minimize the system energy.  From this
point of view, a shift in Eq.~\eqref{eq:shift} implies that the vacuum
should be reorganized to result in an additional mass;
$M^2=m_0^2 + \kappa^2 > m_0^2$.  Because the dynamical mass originates
from a condensate of fermion and anti-fermion (i.e.\
$\sim \langle\bar{\psi}\psi\rangle$), we can rephrase our finding as
an enhanced condensate by finite $\kappa$, and this can be physically
interpreted as a novel realization of the Magnetic Catalysis assisted
by spatial modulation.  See
  Ref.~\cite{PhysRevLett.113.091102,*PhysRevLett.114.181601} for
  analogous discussions.
\vspace{0.3em}

\paragraph*{Spatially Assisted Chiral Magnetic Effect:}

A clean experimental environment for the detection of the chiral
magnetic effect is a Dirac semimetal in a parallel $\bE$ and
$\bB$~\cite{PhysRevLett.113.027603,*Li:2014bha}.  Non-zero chirality
is generated by $\bE\cdot\bB\neq 0$ according to $w$ in
Eq.~\eqref{eq:w_EB}, which together with $\bB$ produces a
topological current $\propto w B \sim EB^2$ with an
  exponential suppression factor; see
  Ref.~\cite{PhysRevLett.104.212001,*Fukushima:2015tza} for an
  explicit form.  The induced current is
balanced between the production rate $w$ and the relaxation time
$\tau$, and thus a dynamically and spatially assisted $w$ would
increase the balanced value of the topological current by
  an exponential factor with the residual Dirac mass replaced with the
  shifted one according to Eq.~\eqref{eq:shift}, which should
be advantageous for more precise measurements of the
  electric conductivity. 
\vspace{0.3em}

\paragraph*{Conclusions:}

The Schwinger Mechanism and the Magnetic Catalysis were explored in
parallel electromagnetic fields with dynamically modulated electric
and spatially modulated magnetic perturbations.  We found that not
only was the pair production rate enhanced with the dynamically
assisting electric field but also with the spatially assisting
magnetic field.  The former effectively reduces the particle mass $m$
multiplicatively, while we found that the latter reduces $m$
subtractively.  For electromagnetic fields smaller
than $m^2$, the dynamically assisting effect is significant, whereas
for electromagnetic fields comparable or larger than $m^2$ as could be
manifested in the high-energy nuclear experiment and/or in the
table-top experiment with massless Dirac dispersions, the spatially
assisting effect that we discovered is dominating.  Our finding of the
effective mass shift due to positive curvature magnetic
fields is robust
regardless of physical processes, so that we can apply it to a static
property of the vacuum.  That is, the dynamical mass for interacting
fermions should be also shifted accordingly, and we discussed that
spatially increasing magnetic fields should increase the
dynamical mass or the condensate.

We can anticipate an intriguing application of the Spatially Assisted
Schwinger Mechanism to lower the critical field strength in high-power
laser facilities (in the hard X-ray region) as well as ion-laser
collisions~\cite{PhysRevLett.103.170403}.
More so, it would be of paramount interest to see whether a field
configuration which permits $\kappa^2>m^2$ is easily realized for Dirac/Weyl
semimetals in appropriate optical environments.
Before the vacuum reorganization is complete which takes a finite
time, our results suggest that the particle production on a wrong
vacuum with $\kappa^2>m^2$ explodes transiently, and such a
transitional behavior might be related to some magnetically driven
instabilities such as the Weibel and the chiral plasma
instability~\cite{Akamatsu:2013pjd}.

It is of further interest to study what effects a perpendicular
spatial inhomogeneity and/or a temporal modulation in the magnetic
field may have.  It is
known that a spatially modulated electric field inhibits the pair
production~\cite{Nikishov1970346}.
Also, for constant magnetic fields perpendicular to the electric field,
it was found the pair production
decreases~\cite{PhysRevA.86.013422}.  These questions deserve further
investigations in the future.

\begin{acknowledgments}
We are grateful to Sanjin~Beni\'{c}, Francois Gelis, Pablo~Morales,
and Igor~Shovkovy for useful discussions.  K.~F.\ was partially
supported by JSPS KAKENHI Grant No.\ 15H03652 and 15K13479.
\end{acknowledgments}

\bibliographystyle{apsrev4-1.bst}
\bibliography{references}

\end{document}